\documentclass[journal ]{new-aiaa}
\usepackage[utf8]{inputenc}
\usepackage{textcomp}

\usepackage[version=4]{mhchem}
\usepackage{graphicx}
\usepackage{bm}
\usepackage{amsmath}
\usepackage{supertabular}
\usepackage{longtable,tabularx}
\usepackage{booktabs}

\usepackage{caption}
\usepackage{comment}
\usepackage{subcaption}
\usepackage{tabularx}
\usepackage{longtable}
\usepackage{color}
\usepackage{colortbl}
\usepackage{mathtools}
\usepackage{relsize}
\usepackage{amsfonts}
\usepackage{url}
\usepackage{multicol}
\usepackage{multirow}
\usepackage{placeins}
\usepackage[printonlyused]{acronym}
\usepackage{rotating}
\usepackage{epstopdf}
\usepackage{enumitem} 
\usepackage{mdwlist} 
\usepackage{afterpage} 
\usepackage{siunitx}
\usepackage{pgfplots}
\pgfplotsset{compat=newest}
\usetikzlibrary{plotmarks}

\usepackage{tikz}
\usepackage{tikz-3dplot}
\usetikzlibrary{calc}
\usetikzlibrary{arrows}
\usetikzlibrary{3d}
\usetikzlibrary{positioning}
\usetikzlibrary{backgrounds}
\tikzset{>=latex}
\tikzset{every picture/.style={font issue=\small},
	font issue/.style={execute at begin picture={#1\selectfont}}
}

\usetikzlibrary{calc,external,patterns,decorations.pathmorphing,decorations.markings}
\usepgfplotslibrary{external}
\newenvironment{myfont}{\fontfamily{pcr}\selectfont}{\par}
\DeclareTextFontCommand{\textcourier}{\myfont}

\usepackage{array}
\newcolumntype{L}[1]{>{\raggedright\let\newline\\\arraybackslash\hspace{0pt}}m{#1}}
\newcolumntype{C}[1]{>{\centering\let\newline\\\arraybackslash\hspace{0pt}}m{#1}}
\newcolumntype{R}[1]{>{\raggedleft\let\newline\\\arraybackslash\hspace{0pt}}m{#1}}

\title{Minimum Radiative Heat and Propellant Aerocapture Guidance with Attitude Kinematics Constraints\footnote{Manuscript accepted for publication in the Journal of Guidance, Control and Dynamics on May 20, 2025. An initial version of this paper was presented at the 2018 AIAA Guidance, Navigation, and Control Conference, 8-12 January 2018, Kissimmee, Florida. Paper number AIAA 2018-1319.}}

\author{Enrico M. Zucchelli \footnote{Postdoctoral Associate, Aeronautics and Astronautics, 77 Massachusetts Avenue; enricoz@mit.edu, AIAA Member.}}
\affil{Massachusetts Institute of Technology, Cambridge, MA, 02139}
\author{Erwin Mooij\footnote{Associate Professor, Aerospace Engineering, Kluyverweg 1,  Associate Fellow AIAA.}}
\affil{Delft University of Technology, Delft, 2629 HS, The Netherlands}

\newcommand{\degree}{$^\circ$}
\newcommand{\be}{\begin{equation}}
\newcommand{\ee}{\end{equation}}

\begin{document}

\maketitle

\begin{abstract}
{Aerocapture leverages atmospheric drag to convert a spacecraft's hyperbolic trajectory into a bound orbit.} 
For some aerocapture missions, heating due to the radiation of high temperature gases in the shock-layer {can be} much larger than the heat due to convection. 
This paper provides analytical proof and numerical validation that radiative heat {load} is minimized by the same trajectory that minimizes the final $\Delta V$: a single switch bang-bang trajectory, starting with lift up. {The proof is very general and is valid for several formulations of radiative heat flux; further, the same proof can be used to conclude that convective heat load, computed according to many of the available formulations, is instead maximized by that trajectory.}
Further, a novel guidance that plans a bang-bang trajectory with constraints in the attitude kinematics is introduced. While achieving performance {similar to that of} the current state-of-the-art, the inclusion of constraints in attitude kinematics allows for much less tuning. Finally, a lateral guidance that makes use of information on the final inclination of the predicted trajectory is introduced. Such guidance allows for very high accuracy in the inclination requirements with only two reversals, by requiring a single parameter to be tuned.
\end{abstract}

\section*{Nomenclature}

\begin{tabbing}
	XXXXX \= \kill
	$ a$   \> orbital semi-major axis, m \\
	$ C_D$	  \> drag coefficient \\
	$ C_L$	  \> lift coefficient \\
	$ D$	  \> drag, N \\
	$ e$   \> orbital eccentricity \\
	$ g_{\delta}$   \> latitudinal component of the gravity, m/s$^2$ \\
	$ g_{r}$   \> radial component of the gravity, m/s$^2$ \\
	$ i$   \> orbital inclination, rad \\
	$L$		  \> lift, N \\
	$m$		  \> mass, kg \\
	$\dot{q}$ \> heat flux at stagnation point, W/m$^2$ \\
	$ Q$   \> integrated heat load, J/m$^2$ \\
	$r$   \> radial distance, \si{\meter} \\
	$R_e$   \> equatorial radius of the Earth, m \\
	$R_N$   \> nose radius, m \\
	$t$   \> time, s \\
	$V$   \> relative speed, m/s \\
	$\gamma$  \> relative flight-path angle, rad \\
	$\delta$ \> latitude, rad \\
	{$\theta$} \> longitude, rad \\
	$\lambda$  \> co-state\\
	$\mu$ \> gravitational parameter of the Earth, m$^3$/s$^2$\\
	$\rho$  \> atmospheric density, kg/m$^3$\\
	$\sigma$ \> bank angle, rad \\
	$\chi$ \> heading, rad \\
	$\omega_{cb}$ \> rotational rate of the Earth, rad/s \\
	\\
	\textit{Subscripts}
	\\
	$_{0}$         \> initial conditions\\
	$_{a}$         \> apoapsis \\
	$_{p}$         \> periapsis \\
	\\
	\textit{Superscripts}
	\\
	$\star$         \> target\\

\end{tabbing}

\section{Introduction}

\begin{figure}[htb]
        \centering
        \includegraphics[width =3.25in]{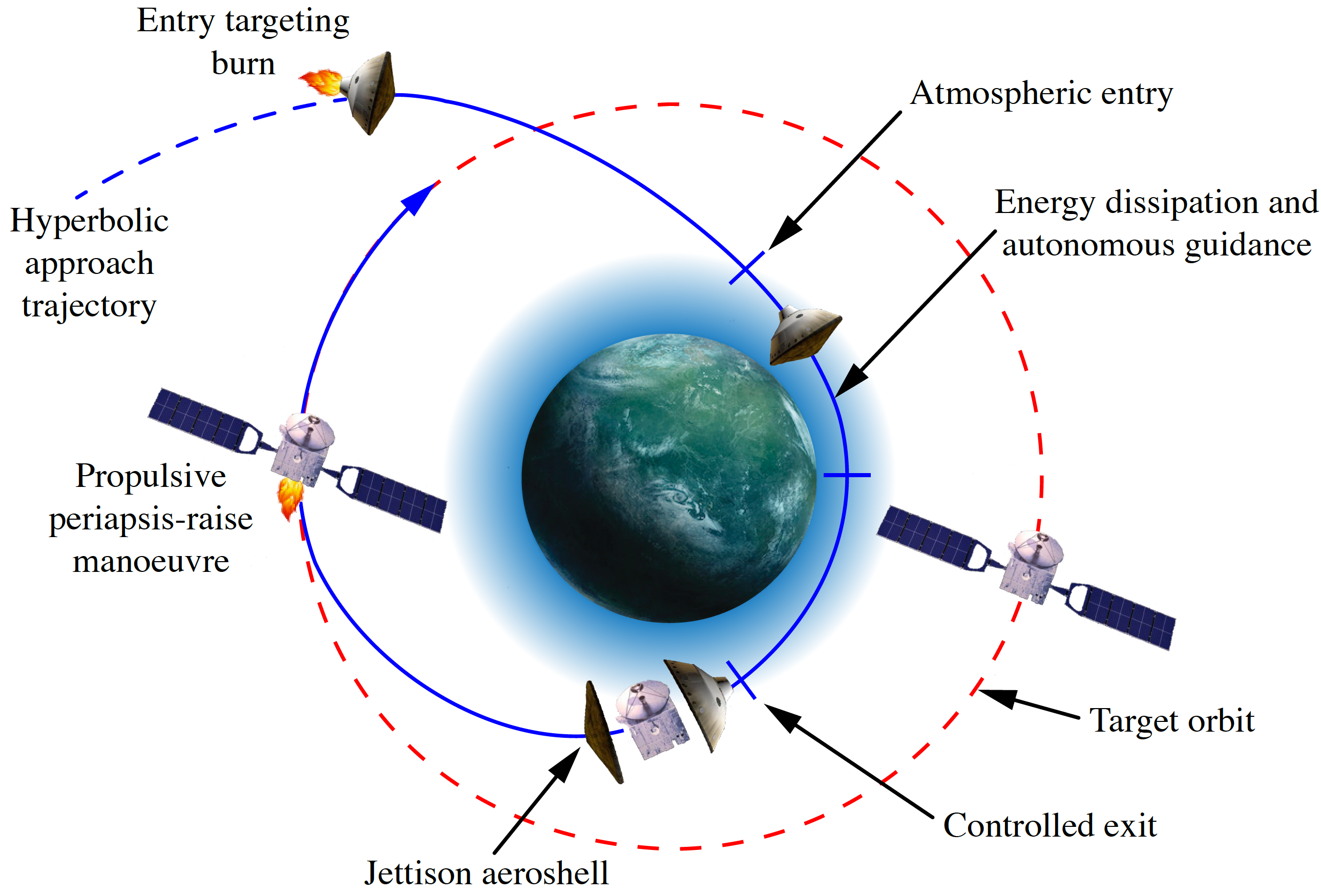}
    \caption{Phases of an aerocapture maneuver.}
    \label{fig:aerocapture}
\end{figure}

Aerocapture, first introduced by Cruz~\cite{Cruz1979}, is an a{ero-assist} maneuver that can greatly facilitate missions to atmospher{e-bearing} celestial bodies~\cite{Hall2005}. Aerocapture achieves orbit insertion from a hyperbolic trajectory, as depicted in Fig.~\ref{fig:aerocapture}. By diving into the atmosphere, the energy of the spacecraft is reduced in a controlled way through the dissipative action of drag. {During this phase, the aeroshell endures intense thermal loads, absorbing and dissipating the heat generated by atmospheric friction}. After the desired amount of energy has been lost, the spacecraft exits the atmosphere{, jettisons the aeroshell,} and coasts to apoapsis. A small propulsive burn is then required to raise the periapsis above the region where drag is still significant. A second small burn is later performed to correct any errors in the target apoapsis altitude. 

{To date,} an aerocapture maneuver has never been attempted, despite the fact that most of the technology is considered ready by several studies~\cite{Munk2008,spilker,girija2022}.
Interest in aerocapture has been growing lately as it would be an enhancing technology for mission to the ice giant planets, Neptune and Uranus~\cite{Saikia,Girija,DESHMUKH2020375}; specifically, attractiveness comes from the fact that it would allow faster, higher energy trajectories, since less propellant mass would be required to decelerate {with aerocapture compared to propulsive insertion}. The mass reduction granted by decreased propellant consumption is partially offset by the requirement of a heat shield to protect the spacecraft {during the atmospheric phase}. Because of the extremely high temperatures encountered during this class of maneuvers, the heat shield is generally ablative, meaning that the mass increases for larger total heat load. Hence, an optimal aerocapture should not only minimize the $\Delta V$ of the post-maneuver burns, but strike a balance between total heat load and total propellant consumption. {Since mission success and risk minimization are the priority, trajectory optimization is however conditioned on all safety requirements being satisfied first.}
The minimum $\Delta V$ aerocapture follows a bang-bang trajectory, beginning with a full lift up command, and with a single switch~\cite{Miele1990,Lu2015}. The trajectory minimizing the $\Delta V$ also minimizes both the heat flux peak and the peak dynamic pressure~\cite{Miele1992}. However such a trajectory {\it maximizes} the integrated convective heat {load}, which is in turn minimized by a bang-bang trajectory that begins with a full lift down command~\cite{Sigal2001}.

A large source of heat during atmospheric missions is due to the radiation of high-temperature gases in the shock layer~\cite{Tauber1991,brykina2019approximation}. For lunar return aerocapture, which is the lowest speed aerocapture possible on Earth, heat load due to convection and heat load due to radiation are comparable in magnitude for trajectories flown by Orion~\cite{confpaper}. At higher {entry }velocities, {radiative heat transfer becomes more significant, eventually dominating convective heat transfer at higher speeds due to its generally higher velocity exponent in the formulation.}
The density, composition, and scale height of the atmosphere also affect the ratio between the two {modes of heat transfers; for entry vehicles with diameters ranging between 1 and 4~m}, radiative heat {load} is generally negligible for missions to Uranus~\cite{morgan2024thermal}, may be relevant on Mars {and Neptune }depending on entry velocity~\cite{grinstead2009shock,hollis2004preliminary,scoggins2025forebody}, and is instead expected to be a major source of heat {load} for missions to Titan {and Jupiter}~\cite{olejniczak2003analysis,Wright2005coupling,brandis2017titan,ritter2006jupiter}{, albeit to date aerocapture at Jupiter is still considered infeasible}~\cite{girija2022}. {As the nose radius increases, radiative heat flux rises while convective heat flux decreases, making radiation more prevailing} for larger spacecraft.
The first key contribution of this paper is to analytically prove, given a few assumptions, that the minimum radiative heat load aerocapture trajectory coincides with the trajectory minimizing the $\Delta V$. The proof encompasses many of the several empirical radiative heat {flux} formulations that have been proposed in the last 40~years.
The proof is validated numerically for a variety of conditions and formulations of radiative heat flux. The same proof can be used to deduce the fact that convective heat {load} is instead maximized by that trajectory, as previously shown by Sigal and Guelman~\cite{Sigal2001}; however, this proof generalizes to more empirical formulations of convective heat flux than the one considered by Sigal and Guelman. Finally, for cases where the heat {load} due to convection is negligible compared to radiative heat {load}, the same trajectory can minimize $\Delta V$ and total heat load at the same time, avoiding the need to seek trade-offs. 

To fly the minimum {radiative }heat {load}, minimum $\Delta V$ aerocapture, this paper introduces a novel optimal numerical predictor-corrector~(NPC) guidance. Initial efforts to develop an optimal guidance for aeroassisted maneuvers focused on orbital plane change with minimum energy loss~\cite{McEneaney,Calise1,Calise2}. Early aerocapture guidance schemes were derived from the Apollo skip entry guidance~\cite{gurley} which relies on the tracking of a nominal trajectory. Aerocapture NPCs seek a constant bank angle to be kept throughout the entirety of the trajectory~\cite{powell} such that the desired apoapsis is targeted. 
Optimal aerocapture guidance was initially proposed to minimize the control effort~\cite{McEneaney}. 
Optimality for closed-loop guidance in terms of $\Delta V$ for an aeroassisted maneuver was later considered by Evans and Dukeman for aerobraking~\cite{Evans1995}. The minimum $\Delta V$ is obtained by maximizing exit velocity for a given apoapsis. To this end, the algorithm is divided into two phases. In the first phase, the algorithm integrates the trajectory with a constant, quasi full lift down bank angle. If the predicted apoapsis is lower than the target one, full lift up is commanded; else, Phase~2 is triggered, in which the algorithm then behaves like the guidance in Ref.~\cite{powell}. This approach achieves optimality avoiding online trajectory optimization. A bank angle margin is required, which decreases performance, but reduces the risk of skipping out. Fully Numerical Predictor-corrector Aerocapture Guidance~(FNPAG)~\cite{Lu2015} for aerocapture {improves on the previous approach} by tuning the optimal bank angle margins as a function of entry angle and velocity. The parameter is tuned for a two-dimensional grid of different entry conditions; for each point of the grid, Monte Carlo runs are required for the tuning. At the beginning of the maneuver, the parameter is interpolated from the grid. Recent developments have been considering uncertainties directly in the planning via stochastic and robust optimization under uncertainty, similarly to stochastic model predictive control~\cite{Heidrich,9172724}, or with two stage optimization under uncertainty~\cite{zucchelli}, for cases where perturbations are so large that there is no control profile that leads to feasible trajectory under all perturbations. Whilst these approaches are more robust than the FNPAG, they are also more computationally demanding. On the other hand, common alternatives to NPCs are analytical predictor-correctors~(APCs)~\cite{Masciarelli,Hamel,cihan,chen}, which make more simplifications than NPCs, and result in a faster, but potentially less robust, guidance algorithm. 

This paper improves on the FNPAG by showing that the major reason why different bank angle margins are needed for different entry conditions is the unconstrained kinematics in the motion planning during Phase~1. Taking the kinematics into account makes the planning more robust and {the guidance} easier to tune. The rotation from the bank angle of Phase~1 to the bank angle of Phase~2 occurs instantaneously in the planning of the guidance logics of Refs.~\cite{Lu2015} and \cite{Evans1995}. In fact, for the majority of re-entry guidance systems, attitude kinematics are not taken into account while planning, and their effect is usually negligible. 
A rotation from full lift up to full lift down may take {up to} 15 seconds, which is short relatively to the entirety of an atmospheric re-entry flight. 
During aerocapture however, 15 seconds are enough to dissipate more than 30\% of the difference in energy between initial and final states~\cite{confpaper}. The percentage is strongly dependent on the initial flight path angle and velocity, which affect both the overall duration of the maneuver and the switching time. Therefore, a predictor-corrector guidance for aerocapture greatly benefit{s} by including the effects of this rotation in the prediction. The guidance proposed in this paper takes attitude kinematics constraints into account. As a result, it performs similarly to that of Ref.~\cite{Lu2015}, but using the minimal tuning of Ref.~\cite{Evans1995}.

The proposed longitudinal guidance is complemented by a novel robust lateral guidance. The lateral guidance aims to minimize the number of bank reversals. It does so by predicting the trajectory as if a bank reversal was happening immediately, and, similarly to the longitudinal guidance, it includes attitude kinematics constraints. This allows the lateral guidance to take into account several perturbations induced by the reversal, such as the finite time of the bank reversal, and the effect of accelerations that depend on the heading or on the position of the spacecraft, such as Coriolis. The proposed lateral guidance has been partly tailored to the longitudinal guidance proposed in this paper, but is easily applicable to other entry problems.

The contributions of this paper can be summarized as follows:
\begin{itemize}
	\item The analytical proof that the minimum radiative heat load aerocapture is the same trajectory that leads to a minimum $\Delta V$. For very high-speed aerocapture, convective heat flux is much smaller than radiative heat flux; thus, that same trajectory minimizing heat load minimizes the total integrated heat load as well, in addition to minimizing $\Delta V$.
	\item The introduction of the optimal aerocapture guidance with attitude kinematics constraints~(OAK), an optimal NPC guidance for longitudinal aerocapture guidance that achieves optimal results with minimal tuning by constraining the attitude kinematics in the planning.
	\item The introduction of a lateral guidance that assumes immediate start of a bank reversal, allowing for almost single reversal lateral control. 
\end{itemize}

This paper continues as follows. Section~\ref{sec:optimal_aero} describes the dynamics of aerocapture. Section~\ref{sub:min_rad_heatload_aero} provides the analytical proof of the form of the trajectory that minimizes radiative heat {load}. Section \ref{sec:guidance} introduces both the longitudinal and the lateral logic of the novel guidance of this paper. Section \ref{sec:results} shows the results obtained from extensive simulation campaign. The guidance is tested for different entry conditions and vehicles. Section \ref{sec:conclusions} concludes this paper.


\section{Aerocapture Flight Dynamics}
\label{sec:optimal_aero}

The dynamics of a vehicle undergoing aerocapture are dominated by gravitational and aerodynamic forces. Including the $J_2$ component of the gravity field, the equations of motion of a vehicle in the atmosphere of a planet are \cite{Miele1989}:
\begin{align}
\label{eq:eqmV} 
\dot{V}=&-\frac{D}{m}-g_r\sin\gamma-g_{\delta}\cos\gamma\cos\chi+\omega_{cb}^2r\cos\delta(\sin\gamma\cos\delta-\cos\gamma\sin\delta\cos\chi)\\
V\dot{\gamma}=&\frac{L\cos\sigma}{m}-g_r\cos\gamma+g_{\delta}\sin\gamma\cos\chi+2\omega_{cb}V\cos\delta\sin\chi+\frac{V^2}{r}\cos\gamma+\\
\nonumber+&\omega_{cb}^2r\cos\delta(\cos\gamma\cos\delta-\sin\gamma\sin\delta\cos\chi)\\
V\cos\gamma\dot{\chi}=&\frac{L\sin\sigma}{m}+g_{\delta}\sin\chi+2\omega_{cb}V(\cos\gamma\sin\delta-\sin\gamma\cos\delta\cos\chi)+\\
\nonumber+&\frac{V^2}{r}\cos^2\gamma\tan\delta\sin\chi+\omega_{cb}^2r\cos\delta\sin\delta\sin\chi\\
\dot{r}=&V\sin\gamma\\
\dot{{\theta}}=&\frac{V\sin\chi\cos\gamma}{r\cos\delta}\\
\label{eq:atm_equations_spher_relv}
\dot{\delta}=&\frac{V\cos\gamma\cos\chi}{r}
\end{align}
where $V$ is the relative velocity, $\gamma$ is the relative flight-path angle, $\chi$ is the relative heading angle, $r$ is the radial distance from the center of the planet, and ${\theta}$ and $\delta$ are the longitude and latitude, respectively. $L$ and $D$ are the aerodynamic lift and drag, $m$ is the vehicle mass, $\omega_{cb}$ is the planet angular velocity, and $\sigma$ is the bank angle. $g_r$ and $g_\delta$ are the two components of the gravity field, when the $J_2$ zonal term is included:
\begin{equation}
g_\delta =-\frac{3}{2}\mu J_2 \frac{R_e^2}{r^4} \sin 2\delta
\end{equation}
\begin{equation}
g_r = \mu\left[-\frac{1}{r^2}+\frac{3}{2} J_2 \frac{R_e^2}{r^4}\left(3\sin^2\delta-1\right)\right]
\end{equation}
Standard coordinate transformations can link the above model to Keplerian orbits. Given a target circular orbit with semi-major axis $a^\star$, and assuming an exit orbit with semi-major axis $a$ and apoapsis $r_a=a(1+e)$, the magnitude of planar $\Delta V$, required to raise the periapsis, as well as to correct the apoapsis, is~\cite{Lu2015}:
\begin{equation}
	\Delta V = \left\|\Delta V_{1}\right\|+\left\|\Delta V_{2}\right\|=\sqrt{2\mu}\left(\left\|\sqrt{\frac{1}{r_a}-\frac{1}{r_a+a^\star}}-\sqrt{\frac{1}{r_a}-\frac{1}{2a^\star}}\right\|+\left\|\sqrt{\frac{1}{2a^\star}}-\sqrt{\frac{1}{a^\star}-\frac{1}{r_a-a^\star}}\right\|\right)
\end{equation}
This equation can easily be generalized to elliptical target orbits. However, the {elliptical} case will not be considered here: {circular target orbits benefit the most from aerocapture, since, for same periapsis radius, they are the most expensive to obtain with propulsive capture.}

The $\Delta V$ needed because of a change in inclination is:

\begin{equation}
	\Delta V_{i}\approx 2V\sin(\frac{\Delta i}{2})
\end{equation}
Finally, the total $\Delta V_{tot}$ that includes in-plane and out-of-plane components, is:

\begin{equation}
	\Delta V_{tot}=\sqrt{\Delta V_{1}^2+\Delta V_{i}^2}+\Delta V_{2}
\end{equation}

The out-of-plane correction is assumed here to be occurring entirely during the first burn. While the optimal strategy would be to leave a small portion of the correction for the second burn, the difference is negligible.

\section{Minimum Radiative Heat Load Aerocapture}
\label{sub:min_rad_heatload_aero}
It is possible to infer some analytical conclusions on the optimal aerocapture trajectory assuming that the central body is not rotating, and that there are no requirements on the final inclination nor final argument of periapsis.
In such a case, an aerocapture leads to a minimum $\Delta V_{tot}$ if the bank angle trajectory is full lift up, followed by full lift down~\cite{Lu2015}. Peak convective heat flux and structural load are too minimized by a bang-bang full lift up, full lift down trajectory~\cite{Miele1992}. Conversely, an aerocapture leads to minimum total convective heat {load} if the bank angle history is full lift down, full lift up~\cite{Sigal2001}. The objectives thus lead to opposite trajectories. Nonetheless, during aerocapture a major source of heat flux comes from the radiation of incandescent gases in the shock-layer. For lunar return conditions on Earth, the total radiative heat load is comparable to the total convective heat load. For higher velocities, the ratio changes in favor of the radiative heat load. {Depending on entry conditions and spacecraft size, radiation can be a major heat source for aerocapture on several other bodies, such as Mars, Neptune, Titan, and Jupiter}~\cite{grinstead2009shock,hollis2004preliminary,scoggins2025forebody,olejniczak2003analysis,Wright2005coupling,brandis2017titan,ritter2006jupiter}.
Thus, minimization of radiative heat load becomes much more important than minimization of the convective heat load for high speed aerocapture flown by large vehicles. The objective of this is section to prove that the integral of a class of monomial functions of density and velocity is minimized by a bang-bang trajectory. {It is shown that, f}or many empirical formulations of the radiative heat flux\cite{Martin1966,Tauber1991,RUMYNSKII1974173,brandis2014characterization,suttles1974curve}, such a trajectory is full lift up, full lift down, which corresponds to minimizing $\Delta V_{tot}$, the heat flux peak, and the dynamic pressure peak.

Let the heat flux be described as a generic function $f=f\left(\rho,V\right)$, where the density $\rho$ is a a strictly monotonically decreasing function of the altitude $h$. Neglecting requirements on the target orbit plane, for a spherical, non-rotating planet, Eqs. \eqref{eq:eqmV}-\eqref{eq:atm_equations_spher_relv} reduce to:
\begin{align}
	\label{eq:Vdot}
    	\dot{V}=&-\frac{D}{m}-\frac{\mu}{r^2}\sin\gamma\\
	V\dot{\gamma}=&\frac{L\cos\sigma}{m}-\frac{\mu}{r^2}\cos\gamma+\frac{V^2}{r}\cos\gamma\\
	\dot{r}=&V\sin\gamma
	\label{eq:rdot}
\end{align}
The cost function to be minimized is:
\begin{equation}
	J=\int_{t_0}^{t_f}f\left(\rho,V\right)
\end{equation}
Let us now assume that $f\left(\rho,V\right)$ is separable, such that it can be decomposed as the product of a function of the density $f_\rho(\rho)$ and a function of the velocity $f_V(V)$. For the special case where $f_\rho(\rho)=\rho$ and $f_V(V) = V^{n_V}$, and assuming $D/m\gg g|\sin\gamma|$, which is valid throughout the most relevant parts of the trajectory:
\begin{equation}
    \int_0^{t_f}  \rho V^{n_V} dt = \int_0^{t_f}  \frac{\rho V^{n_V}}{\dot{V}}\frac{dV}{dt} dt = - m\int_{V_0}^{V_f}  \frac{\rho V^{n_V}}{D}dV = - \frac{2m}{S C_D} \int_{V_0}^{V_f} {V^{n_V-2}}dV = \frac{2m}{(n_{{V}}-1)S C_D}\left(V^{n_V-1}_0-V^{n_V-1}_f\right).
\end{equation}
The lift up-lift down sequence provides the highest final velocity, and thus minimizes the cost integral as long as $n_V>1$. However, the relative difference is negligible, especially for high $n_V$, where the initial velocity to the $n_V^{\text{th}}$ power would dominate over any differences in the final velocity. 
This result can be generalized to any positive function $f(\rho,V)$ of the form $\rho f_V(V)$, where the dependency on the velocity $f_V(V)$ is an arbitrary function, as long as it can be separated from~$\rho$:
\begin{equation}
    \int_0^{t_f}  \rho f_V(V) dt = - \frac{2m}{S C_D} \int_{V_0}^{V_f} \frac{f_V(V)}{V^2}dV = \frac{2m}{S\, C_D}\left(g(V_0)-g(V_f)\right),
\end{equation}
where
\begin{equation}
    g(V)=\int \frac{f_v(V)}{V^2}dV,
\end{equation}
and only depends on the exit value of $V$ which, again, has negligible dependency on the trajectory.

Consider now a more general case where $f_\rho(\rho)=\rho^{m_{\rho}}$.
For a given trajectory, one has that the velocity continuously decreases as long as $D/m>V\sin\gamma$. Therefore, the density $\rho(V)$ can be parameterized as a unique function of $V$. Neglecting the dependency of the final velocity $V_f$ on the trajectory, which is small, the previous result shows that
\begin{equation}
\label{eq:constantintegratedtrajectory}
    \int_{t_0}^{t_f} \rho(V) f_V(V) dt = C,\qquad \forall\, \rho(V),\,f_V(V),
\end{equation}
where $C$ is a constant independent of the flown trajectory.
The following equality
\begin{equation}
    \int_{t_0}^{t_f}  \rho^{m_{\rho}}(V) f_V(V) dt = \int_{t_0}^{t_f}  \rho^{m_{\rho}-1}(V)\times \left(\rho(V) f_V(V)\right) dt,
\end{equation}
where the $\times$ symbol has been used only to stress the separation in two factors, leads to realizing that the integral on the left side 
can be seen as the previous integral of Eq.~\eqref{eq:constantintegratedtrajectory}, where each infinitesimal $\rho(V) f_V\left(V\right) dt$ is multiplied by $\rho^{m_{\rho}-1}$. Thus, if it exists, the optimal trajectory is such that
\begin{equation}
    \rho_{\text{opt}}^{m_{\rho}-1}(V) < \rho^{m_{\rho}-1}(V)\qquad \forall\, V.
\end{equation}
For $m_{\rho}>1$, that implies that the density has to be minimized, and therefore the altitude maximized, for \textit{every} value of the velocity. Such trajectory is flown by first flying lift up, then lift down. {An explanation for why that is the case is provided in the Appendix}. For $m_{\rho}<1$ instead, the opposite trajectory should be flown. 
Note that for $f_V(V)=V^3$ and $m_{\rho}=0.5$ one has the analytical equation for convective heat flux, hence this proof includes the minimum convective heat load proof of Ref.~\cite{Sigal2001}, and expands on it by offering an analytical proof for why the minimum convective heat trajectory has only one switch, starting with lift down. Additionally, this proof generalizes to {additional} convective heat {flux formulations} such as the Detra-Hidalgo~\cite{detrahidalgo}, for which the exponent $n_V$ is 3.15 instead of 3.

Let us generalize now the problem {further} to any separable function {of the form }$f(\rho,V)=f_\rho(\rho)f_V(V)${, without limitations on the form of $f_\rho(\rho)$. Similarly as before}:
\begin{equation}
    \int  f_\rho(\rho(V) f_V(V) dt = \int  \rho^{-1}(V)f_\rho(\rho(V)){\times}\left({\rho(V)}f_V(V)\right) dt
\end{equation}
Now, the condition for optimality is:
\begin{equation}
    \frac{f_\rho(\rho_{\text{opt}}(V))}{\rho_{\text{opt}}(V)} < \frac{f_\rho(\rho(V))}{\rho(V)}\qquad \forall\, V.
\end{equation}
If $f_\rho(\rho)$ increases superlinearly for all $\rho\in\mathbb{R}_+$, the optimal trajectory is lift up-lift down; if $f_\rho(\rho)$ increases sublinearly or decreaseas for all $\rho$, the optimal trajectory is lift down lift up; if $f_\rho(\rho)$ is linear the integral cost is independent of the trajectory. For all other cases, \textit{e.g.,} when $\rho^{-1}f_\rho(\rho)$ is not monotonic, no conclusion can be drawn based on this proof alone. Table~\ref{tab:commandorders} summarizes the results of this section.
It is straightforward to expand the proof of this section to any functions of the form
\begin{equation}
    f(\rho,V) = \sum_i f_{\rho,i} (\rho) f_{V,i}(V),
\end{equation}
as long as $f_{\rho,i}$ are all either superlinear or sublinear in $\rho$. 
\begin{table}[]
	\centering
	\caption{Optimal control sequences as a function of $f(\rho,V)$.}
	\label{tab:commandorders}
		\begin{tabular}{ccc} \hline \hline
			 properties of $f(\rho,V)$ & properties of $f_\rho(\rho)$ &    optimal lift sequence    \\
    \hline
            separable & $f_\rho(\rho)$=$\rho$ &  {unresolved} \\
			separable & $f_\rho(\rho)$ superlinear & up-down \\
			separable & $f_\rho(\rho)$ sublinear & down-up \\
            separable & $\rho^{-1}f_\rho(\rho)$ not monotonic & unresolved\\
            not separable & N/A & unresolved\\
             \hline \hline
	\end{tabular}
\end{table}

Many empirical radiative heat flux formulations can be separated between functions of density~$f_\rho(\rho)$ and function of ve{lo}city~$f_V(V)$. Examples include the formulas by Martin ~\cite{Martin1966}, Tauber and Sutton\cite{Tauber1991}, Brandis and Johnston~\cite{brandis2014characterization}, and more~\cite{tauber2012stagnation,RUMYNSKII1974173,suttles1974curve}.
In all mentioned cases $f_\rho(\rho)$ grows superlinearly\footnote{The Tauber-Sutton formula is proportional to $\rho^{1.22}R_n^a$, where $R_n$ is the nose radius and $a \propto \rho^{{-0.325}}$. It is therefore not straightforward to determine for which values it is monotonically increasing when $R_n > 1$ m. However, $a$ is capped between 0 and 1, and therefore non-monotonicity, if any, would be for a small interval. Similar claims can be made about Brandis-Johnston's formula. The formula of Ref.~\cite{suttles1974curve} includes a factor equal to $\rho^{1.2-0.01223V}$, $V$ in km/s, which is not separable; however, the dependency is weak.}, and the optimal trajectory is lift up-lift down. As the formulation{s} for convective heat {flux} {are} instead sublinear in $\rho$, this proof offers no insights on the trajectory minimizing the sum of convective and radiative heat {load}.
\begin{figure}
    \centering
    \includegraphics[trim={0 0.1cm 0 0},clip]{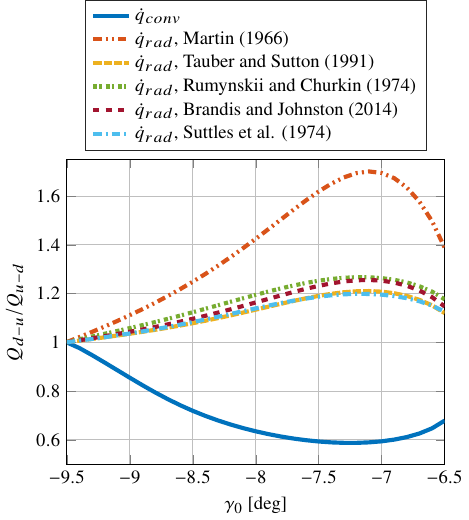}

    \caption{Heat load ratio between down-up and up-down trajectories with different heat flux equations.}
    \label{fig:heatloadratio_formulas}
\end{figure}
Fig.~\ref{fig:heatloadratio_formulas} shows the ratio between heat load flying a down-up sequence, $Q_{d-u}$, and heat load flying an up-down sequence, $Q_{u-d}$, as a function of entry flight path angle for various formulations of radiative heat flux, as well as for the convective heat flux. The trajectories all have initial velocities of 16~km/s and target an apoapsis {altitude} of 500~km. The atmosphere is assumed exponential with scale height of 10~km. The equations of motion used are the simplified set~\eqref{eq:Vdot}-\eqref{eq:rdot}. Radiative heat {load, when computed with} any of the {heat flux formulations in the figure} is lowest for up-down trajectories. At the same time, convective heat load is minimized by a down-up trajectories. The impact of the chosen trajectory on radiative heat {load} highly depends on the empirical formulation of choice, as well as the initial flight-path angle.
\begin{figure}
    \centering
    \includegraphics[trim=60mm 108mm 65mm 110mm, clip,]{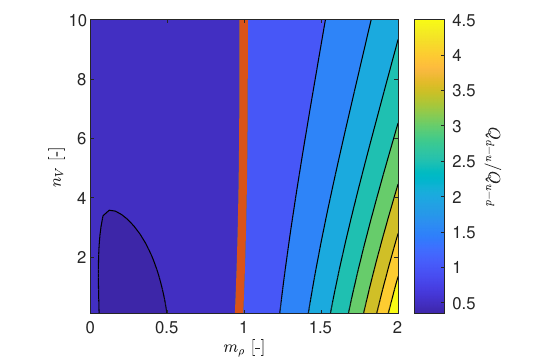}
    \caption{Heat load ratio between down-up and up-down trajectories, varying exponents.}
    \label{fig:heatloadratio_m_n}
\end{figure}
Figure~\ref{fig:heatloadratio_m_n} shows the same ratio for entry flight path angle of $-7^\circ$ for different monomial expressions of $f(\rho,V)=c\rho^{m_{\rho}}V^{n_V}$ with varying exponents. {The red line lies where the ratio is equal to 1}. Increasing $m_{\rho}$ leads to favoring the up-down trajectory more, as expected. On the other hand, higher $n_V$ leads to reduced differences between the flown trajectories. The latter result makes intuitive sense: the higher the exponent on $n_V$, the more relevant the beginning of the trajectory is, when maneuverability is limited, and the difference between flown trajectories is small.

In conclusion, this section prove{s} that during aerocapture the integral of any separable function $f_\rho(\rho) f_V(V)$, where both $f_\rho(\rho)$ and $f_V(V)$ are positive and $f_\rho(\rho)\rho^{-1}$ is monotonic, is minimized by a bang-bang trajectory with a single switch. The order of commands only depends on whether $f_\rho(\rho)$ increases superlinearly {(like for most of the formulations of radiative heat flux)}, in which case the trajectory begins with lift up, or sublinearly {(like for the formulations of convective heat flux)}, in which case the trajectory begins lift down.

\section{Optimal Aerocapture Guidance with Attitude-Kinematics Constraints}
\label{sec:guidance}

A bang-bang trajectory {requires} a long-lasting rotation, from lift up to lift down.
Assuming a maximum angular rate of 15\degree/s and a maximum angular acceleration of 5\degree/s$^2$~\cite{bihari2011orion},
a rotation of {105}\degree~{lasts} around 11~s. In such a time more than 20\% of the total energy difference may be depleted. An additional {challenge} consists of the fact that, depending on the entry conditions, the rotation {may} occur at very different moments of the trajectory.
For a shallow entry, the rotation occurs very soon, when dynamic pressure is small, and thus the error in modeling does not impact the prediction much. For a steeper entry, the rotation would occur later, when the dynamic pressure is larger, causing larger errors. Consequently, we propose a guidance logic that includes such a rotation in the trajectory planning.

This section introduces the Optimal aerocapture guidance with Attitude-Kinematics constraints (OAK), which expands on the FNPAG~\cite{Lu2015} by including a simplified model for the rotation of the vehicle. Similarly to FNPAG, the trajectory is divided into two phases.

\subsection{Rotation Model}
\label{subsec:rot}
\begin{figure}[bt]
    \centering
    \includegraphics[trim={0 0.35cm 0 0},clip]{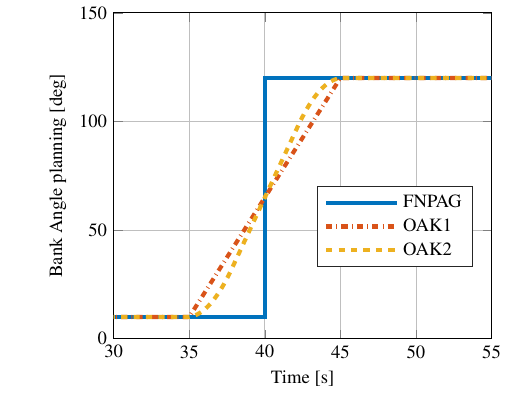}
    \caption{Bank angle profile planned by FNPAG, OAK1, and OAK2.}
    \label{fig:traj_planningsS}
\end{figure}

{Two guidance modes are developed, OAK1, and OAK2. In OAK1, the rotation is modeled as occurring with infinite angular acceleration, but with an average angular rate of $\dot{\sigma}_{exp}=10.5$\degree s$^{-1}$, where the value is the average rotational speed assuming a rotation of {105}\degree ~and accelerations and decelerations at 5\degree s$^{-2}$. The velocity $\dot{\sigma}_{exp}$ can be changed to adapt to different values of the rotations. In OAK2, the acceleration is instead modeled as finite, and the bank angle planning consists of a constant acceleration phase, followed by a period of constant angular velocity (if any), and constant deceleration until $\sigma_d$ is reached. This mode allows the use of the true values of expected angular velocity and acceleration; for this work, the values are, respectively, 15\degree s$^{-1}$ and 5\degree s$^{-2}$. OAK1 is simpler than OAK2 and can be integrated with high-order methods by splitting the integration in only two phases, but requires making some approximations. On the other hand, OAK2 is more complex, and requires splitting the integration into three or four different phases when integrating with high-order methods. At the same time, it is more realistic, and its integration is smoother than OAK1 when integrating the trajectory with a single segment.}
Figure \ref{fig:traj_planningsS} shows the difference between the planning {during}, as in Refs~\cite{Evans1995,Lu2015}, and the bank-angle planning in the proposed algorithm{s, for the case of a rotation from $\sigma_0$ to $\sigma_d$.}

\subsection{Longitudinal Guidance}
As for {FNPAG}, the longitudinal guidance scheme is divided into two phases. {In both cases, during the first phase, the commanded bank value remains constant and equal to a preset angle of $\sigma_0$, which is a low value, close to 10\degree}.
The {main} difference between the proposed guidance and {FNPAG} {is} in how the rotation {is} modeled{: while FNPAG assumes instantaneous change in bank angle between Phase~1 and Phase~2, the OAK schemes use the rotational model described in Sec.~\ref{subsec:rot}.} 
{Another, more subtle difference between the two algorithms is that during Phase~1 OAK does not seek a switching time; instead, it directly calculates the outcome as if a rotation to $\sigma_d$ were initiated immediately.} If the so-predicted apoapsis is lower than the desired apoapsis{, then the command remains to a preset value of $\sigma_0$}. Else, the command becomes equal to $\sigma_d$, and Phase~2 is triggered starting from the next guidance call. This causes a delay in the beginning of Phase~2. Nonetheless, {Phase~1} requires only one iteration per guidance call, {thus} the frequency can be {high, making} the delay {small}. 
{As in FNPAG,} during Phase~2, the logic iterates to find a constant bank angle that leads to the desired apoapsis. Given the strong discontinuities {of the problem, during Phase~2} bisection is used, which leads to an accuracy of 0.05\degree~after 13 iterations.

Density filters are implemented as in Ref.~\cite{Lu2015}. In the inner loop, the modeled lift and drag are multiplied by the scale factors $\tilde{\rho}_L$ and $\tilde{\rho}_D$, respectively. Given $\rho_L$,
\begin{equation}
	\rho_L = L/{\bar{L}}
\end{equation}
where $L$ is the sensed lift, and ${\bar{L}}$ is the lift according to the model, the corresponding scale factor $\tilde{\rho}_L$ is updated at each cycle, applying a low-pass filter:
\begin{equation}
	{\tilde \rho}_L^{\left(n+1\right)}={\tilde \rho}_L^{\left(n\right)}+\left(1-k\right)\left(\rho_L-{\tilde \rho}_L^{\left(n\right)}\right)
\end{equation}
The same applies to ${\tilde \rho}_D$. In this work, a value of $k=0.95$ has been used.

\subsection{Lateral logic}
The lateral logic is specific for this guidance scheme. The guidance is designed to target the final inclination with high accuracy and with at most two reversals. Limiting the number of reversals {is crucial to} reduce the impact of the lateral guidance on the longitudinal performance.

During Phase~1, the bank angle is kept constant.
Once Phase~2 begins, the initial sign of the  bank angle is chosen such that the inclination error is reduced. In this phase, the planning assumes that a bank reversal is initiated immediately, and it is simulated with the same rotational constraints described in the previous section. 

Let $\Delta i_{\text{rev}}$ be the approximate change in inclination that occurs during the bank reversal:
\begin{equation}
	\Delta i_{\text{rev}}=\left\|i_{\text{rev}} - i \right\|
\end{equation}
where $i_{\text{rev}}$ is the predicted final inclination at time $t_{\text{rev}} = 2\min(\pi-\sigma,\sigma)/\dot{\sigma}_{exp}$, and $i$ is the instantaneous inclination at the initiation of the current guidance call. A bank inversion is triggered when all of the following conditions are true:
\begin{enumerate}
	\item the inclination error at the end of the predicted trajectory, $\Delta i_{\text{pred}}$, and the current inclination error, $\Delta i$, have opposite signs: $\Delta i_{\text{pred}} \times \Delta i < 0$
	\item $\Delta i_{\text{pred}}$ is smaller than $i_m$ times the current inclination error $\Delta i$ plus $\Delta i_{\text{rev}}/2$: $\Delta i_{\text{pred}}<i_m\left(\Delta i + \Delta i_{\text{rev}}/2\right)$, and
	\item $\Delta i_{\text{pred}}$ is larger than a maximum allowable inclination error threshold $\Delta i_t$: $\|\Delta i_{\text{pred}}\| > \Delta i_t$.
\end{enumerate}

By doing so, and by setting a margin $i_m$, the number of reversals can be limited to two. Too small a margin may lead to large final errors in the inclination, whereas a too large margin would lead to additional reversals. Since the latter situation is less problematic, when in doubt a larger margin should be preferred to an excessively small one. If a maximum number of reversals is set, $i_m$ is automatically set to 0 before the last reversal. 
Note that the bank reversal only occurs if the conditions for the reversal are met. This leads to a small, constant deviation in the longitudinal guidance; on the other hand, the prediction is much more accurate in case the reversal happens. 

\begin{figure}[tb!]\centering
	\captionsetup{justification=centering,margin=2cm}
    \includegraphics[trim={0 0.3cm 0 0},clip]{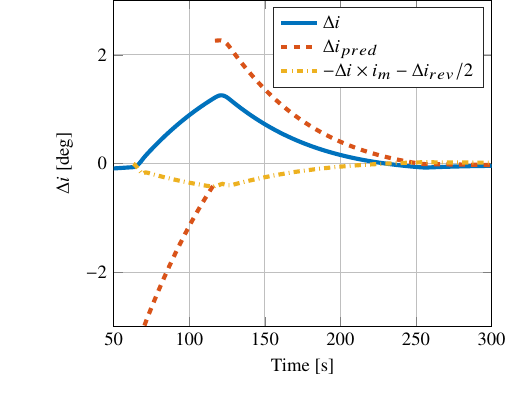}
	\caption{Schematic of the lateral logic during Phase~2; $i_m=0.3$.}
	\label{fig:lateral_logic_10}
\end{figure}

A schematic of the lateral guidance is given in Fig. \ref{fig:lateral_logic_10}. Once the predicted inclination (dashed line) becomes smaller in absolute sense than the sum of current inclination (multiplied by $i_m$) and $\Delta i_{\text{rev}}/2$ (dashdotted line), the reversal begins. The margin $i_m$ is needed because of the many perturbations that may happen after the reversal. In this paper, a 2-reversal strategy with $i_m=0.3$ is used.

\section{Results}
\label{sec:results}

Testing was carried out using a simulator for atmospheric flight on Earth, that was built, verified, and validated. {The baseline atmosphere used is the U.S. Standard Atmosphere of 1976; however,} Earth GRAM large-scale density perturbations \cite{Leslie2011} {are simulated. Additional features} of the simulator include a second-order gravity model, and the Orion {Crew Module~(CM)} and the Apollo {Command and Service Module~(CSM)} databases. The database for the Orion {CM} is for trim conditions at hypersonic flight, {with} uncertainties in aerodynamic coefficients as in Ref.~\cite{Bibb2011}.
The Apollo {CSM} is modeled in trim conditions as in Ref.~\cite{dsmcmoss}; aerodynamic coefficients are uniform{ly distributed} for Apollo, in a range of $\pm20$\% {from the mean}.
A bank angle deadband of 0.1\degree~has been implemented for both vehicles.
The distribution in initial conditions, vehicle parameters, and atmospheric variations are as given in Ref.~\cite{Lu2015}, and reported in Table~\ref{tab:IC}.
The target orbit has an apoapsis altitude of 200~km, and an inclination of 90\degree.
The guidance is triggered once a non-gravitational acceleration larger than 0.05~g is sensed.
It is shut down when the spacecraft crosses 100~km altitude.
The simulation is stopped once apoapsis is reached.
For all simulations, a maximum of two reversals has been chosen, together with an initial margin of 30\%.
\begin{table}[]
	\centering
	\caption{Initial conditions for the lunar return Monte Carlo runs.}
	\label{tab:IC}
		\begin{tabular}{cc} \toprule \toprule
        Parameter & Distribution\\
			 \midrule
            Initial relative velocity $V_0$ [km/s] & $\mathcal{U}(11.05,11.06)$\\
             Entry flight path angle $\gamma_0$~[deg] & $\mathcal{U}(-6.5, -5)$\\
            Initial heading $\chi_0$~[deg] & $\mathcal{U}(-2.1789,-1.1789)$ \\
            Entry interface altitude $h_0$ [km] & 121.9\\
            Initial longitude $\theta_0$~[deg] & 242.75\\
            Initial latitude $\delta_0$~[deg] & -46.67
	\end{tabular}
\end{table}

\subsection{Conceptual comparison}
\begin{figure}[tb!]\centering
	\centering
	\captionsetup{justification=centering,margin=2cm}
	\captionsetup[subfigure]{justification=centering,singlelinecheck=false}
	\begin{subfigure}[t]{0.46\textwidth}\centering
        \includegraphics[trim={0 0.3cm 0 0},clip]{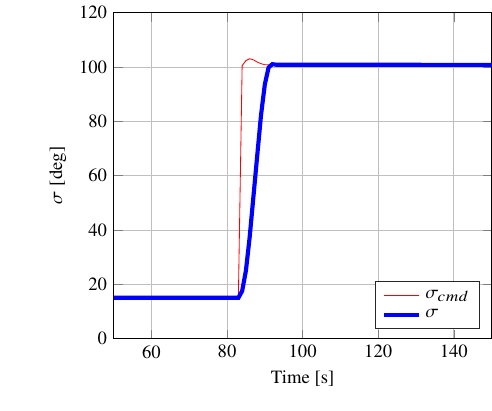}
		\caption{OAK1}
		\label{fig:ver_NPC_mod}
	\end{subfigure}
	\begin{subfigure}[t]{0.46\textwidth}\centering
        \includegraphics[trim={0 0.3cm 0 0},clip]{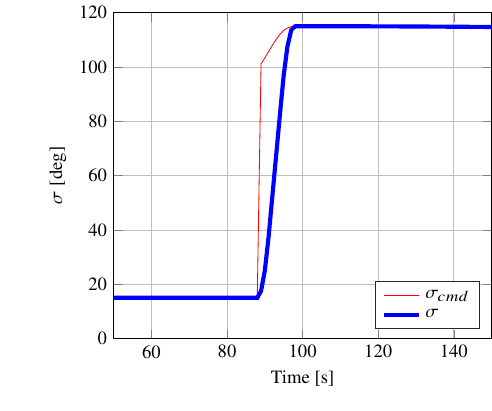}
		\caption{FNPAG}
		\label{fig:ver_NPC_orig}
	\end{subfigure}
	\caption{Bank angle history during flight in ideal conditions, $\sigma_d=$ 100\degree.}
	\label{fig:ver_NPC}
\end{figure}

The behavior of the guidance is {first} evaluated in {a} single, ideal case. {This section aims to be a conceptual comparison between FNPAG and OAK1, in order to validate whether the OAK1 planning is indeed more realistic than the FNPAG.} Here, the environment is modeled exactly the same as in the guidance logic. The bank angle is optimally controlled (according to a minimum-time problem), constrained by maximum angular acceleration and velocity, and is not subjected to any perturbations.
Figure~\ref{fig:ver_NPC} shows the bank angle history under ideal conditions: {$\sigma$ is the actual bank angle, whereas $\sigma_{cmd}$ is the bank angle commanded at that time}. The planned bank angle $\sigma_d$ for Phase~2 is set to~100\degree~ {for both guidance schemes}. {The OAK guidance begins the rotation earlier than FNPAG, and ends the rotation very close to the planned value for $\sigma_d$. On the other hand,} even for {the} ideal conditions {of this case}, with FNPAG the duration of the rotation is long enough to cause a major {unplanned} shift of the final bank angle. {Since the rotation does not occur instantaneously as planned, by the time that the vehicle reaches the value $\sigma_d$, it is too late for that trajectory to be viable any more, and the guidance needs to compensate by commanding an even larger bank angle.} Moreover, such a shift depends on the initial entry angle:
an aerocapture with shallower entry angle lasts longer and is subjected to a smaller maximum dynamic pressure than an aerocapture with a steeper entry angle would be. Consequently, the effect of the bank angle rotation is less pronounced in the former case. This reasoning is in agreement with the optimal tuning {performed} by Ref.~\cite{Lu2015}: in fact, the planned $\sigma_d$ is larger for shallower entries, where the error caused is smaller; a larger margin is instead required for steeper entries.

\subsection{Numeric Comparison}
\begin{figure}[tb!]\centering
	\centering
	\captionsetup{justification=centering,margin=2cm}
	\captionsetup[subfigure]{justification=centering,singlelinecheck=false}
	\begin{subfigure}[t]{0.49\textwidth}\centering
        \includegraphics[trim={0 0.3cm 0 0},clip]{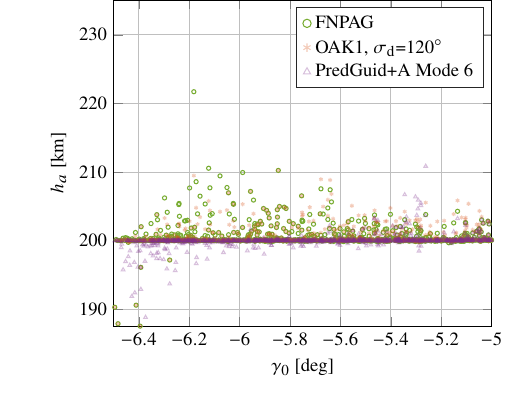}
		\caption{Apoapsis altitude.}
		\label{fig:Lu_vs_sd120_h_apo}
	\end{subfigure}
	\begin{subfigure}[t]{0.49\textwidth}\centering
        \includegraphics[trim={0 0.3cm 0 0},clip]{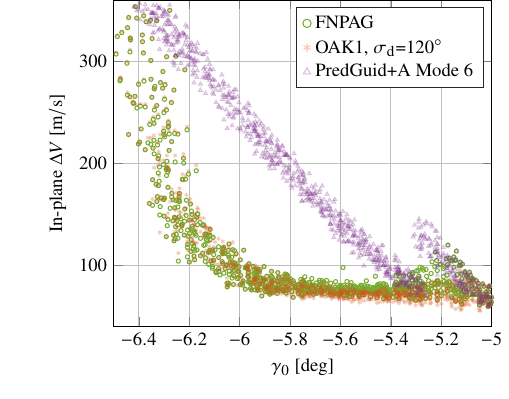}
		\caption{$\Delta V$.}
		\label{fig:Lu_vs_sd120_deltaV}
	\end{subfigure}
	\caption{Numeric comparison between FNPAG, OAK1, and PredGuid+A Mode 6.}
	\label{fig:Lu_vs_sd120}
\end{figure}

{This subsection compares the performance of the OAK1 guidance, $\sigma_d=$ 120\degree, with that of the optimally tuned FNPAG\footnote{The guidance logic {reproduced here differs slightly from the one from Ref.}~\cite{Lu2015}, because {it includes} the lateral logic proposed in this paper.}and of the PredGuid+A Mode 6}~\cite{Lafleur2011}~{(which is equivalent to OAK, with $\sigma_d$ = 0\degree) in} a Monte Carlo campai{gn comprising 1,000 simulations}. {Figure~\ref{fig:Lu_vs_sd120} shows results obtained flying with all three guidance schemes, as a function of initial flight path angle $\gamma_0$. For steep entry angles, several runs end with lower apoapsis than 187.5~km, the lower limit in the plot; however, for all those cases both FNPAG and OAK skip Phase~1, and thus there is no difference between any of the three algorithms in the plot. Similarly, those runs end with $\Delta V$ larger than 350~m/s, and are removed from the right plot.} Figure~\ref{fig:Lu_vs_sd120_h_apo} shows that, in terms of apoapsis accuracy, OAK{1} and FNPAG are rather similar. Similarly, Fig.~\ref{fig:Lu_vs_sd120_deltaV} shows that the OAK1 guidance achieves a {final} $\Delta V$ comparable to that of FNPAG. As expected, the performance of the longitudinal guidance is rather similar between OAK{1} and FNPAG. The main difference {between the two concepts is} the fact that the FNPAG guidance requires extensive tuning, where a value of $\sigma_d$ has to be found via Monte Carlo trials for each combination of entry velocity and angle. On the other hand, the OAK1 guidance can be set by defining a single parameter. In addition, the OAK1 guidance is conceptually more robust, as demonstrated in the previous subsection.
Figure~\ref{fig:Lu_vs_sd120} {also} compares the performance of the OAK1 guidance against PredGuid+A, Mode 6~\cite{Lafleur2011}~(which is equivalent to OAK, with $\sigma_d=$ 0\degree). While the OAK guidance is more accurate in terms of apoapsis targeting, the difference is {small}. On the other hand the OAK guidance performs much better in terms of propellant usage, as expected. For entry angles around -5.8\degree, the $\Delta V$ needed {for periapsis raise after having flown with} PredGuid+A Mode 6 is around 200~m/s, approximately 2.5 times more than {if} OAK {is used instead}. The rightmost branch of the $\Delta V$ with PredGuid+A Mode 6 is a strict consequence of the lateral logic. For the range of entry angles between -6.3\degree~and -5.3\degree, the first bank reversal occurs when the commanded bank angle is smaller than 90\degree. As a consequence, the first rotation is upwards (thanks to the proposed lateral logic being used, rotations after the first one have almost negligible effect on the longitudinal performance). This causes an increase in performance, for the same reason why a full lift up, full lift down trajectory is the optimal one. For shallower entry angles, the opposite happens, causing a decrease in performance. Such pattern does not occur with the OAK guidance. As long as $\sigma_d \geq 105$\degree, the rotation is always downwards. Nonetheless, the rotation is also shorter, and affects the performance less.

\subsection{Sensitivity with respect to $\sigma_d$}
\begin{figure}[tb!]\centering
	\centering
	\captionsetup{justification=centering,margin=2cm}
	\captionsetup[subfigure]{justification=centering,singlelinecheck=false}
	\begin{subfigure}[t]{0.49\textwidth}\centering
        \includegraphics[trim={0 0.3cm 0 0},clip]{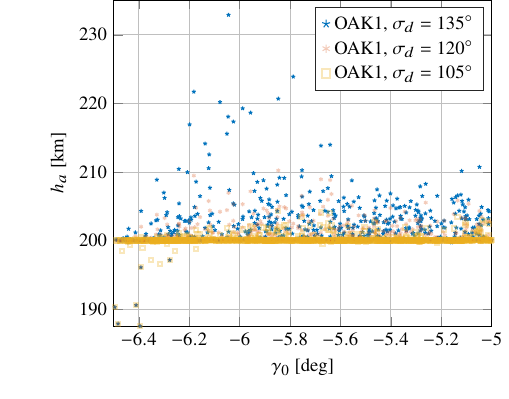}
		\caption{Apoapsis altitude.}
		\label{fig:Orion_r_apo_105_120_135}
	\end{subfigure}
	\begin{subfigure}[t]{0.49\textwidth}\centering
        \includegraphics[trim={0 0.3cm 0 0},clip]{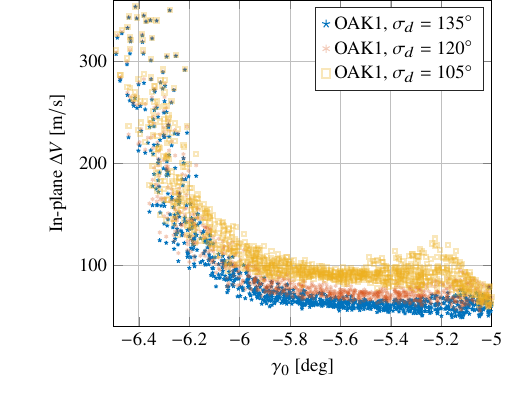}
		\caption{$\Delta V$.}
		\label{fig:Orion_deltaV_105_120_135}
	\end{subfigure}
	\caption{Comparison between different values of $\sigma_d$ for the OAK1 guidance.}
	\label{fig:Orion_105_120_135}
\end{figure}

This subsection analyzes how a change in $\sigma_d$ affects the performance of the guidance over the full range of entry conditions.
Figure \ref{fig:Orion_105_120_135} shows apoapsis altitude and planar $\Delta V$ for the OAK guidance with three values of $\sigma_d$: 105\degree, 120\degree, and 135\degree. While $\sigma_d=135$\degree  ~performs best in terms of planar $\Delta V$, it also causes the largest errors in apoapsis altitude (there is one outlier that is not shown in the figure, for $\sigma_d=$ 135\degree, with $\gamma_0$ = -5.045\degree, and apoapsis altitude 268.7 km). {These cases result from early saturation of the command. While saturation typically does not lead to a significant increase in planar $\Delta V$ unless it occurs too early, it does hinder lateral control.} This, in turn, causes large inclination errors that {require} out-of-plane corrections.

\begin{table}[]
	\centering
	\caption{Summary of OAK1 guidance systems longitudinal performances, for $\gamma_0\in$ (-6\degree, -5\degree).}
	\label{tab:various_sigma_d_long}
		\begin{tabular}{cccccccccc} \toprule \toprule
			&             \multicolumn{4}{c}{In-plane $\Delta V$ [m/s]} & \multicolumn{1}{c}{$\left\|\Delta r_{apo}\right\|$ [km]} & \multicolumn{4}{c}{$\Delta r_{apo}$  [km]} \\ \cmidrule(lr){2-5} \cmidrule(lr){6-6} \cmidrule(lr){7-10}
			$\sigma_d$ [deg] & Mean     & Min      & Max      & Std      & Mean  & Mean   & Min    & Max   & Std  \\ \midrule
			105 &  92.13     & 59.47     & 130.46    & 11.10    & 0.252  & 0.223  & -0.476  & 4.328 & 0.590 \\
			120 &  74.21     & 55.19     & 114.74    & 49.1    & 0.698 & 0.677  & -0.295  & 10.221 & 1.339  \\
			135 &  64.0     & 48.8     & 100.1    & 7.5    & 1.64 & 1.63  & -0.28  & 68.73 & 3.70 \\ \bottomrule \bottomrule
	\end{tabular}
\end{table}

\begin{table}[]
	\centering
	\caption{Summary of OAK1 guidance systems lateral performances, for $\gamma_0\in$ (-6\degree, -5\degree).}
	\label{tab:various_sigma_d_lat}
		\begin{tabular}{cccccccccc}\toprule \toprule
			&               \multicolumn{1}{c}{$\left\|\Delta i\right\|$ [\degree]}     &  \multicolumn{4}{c}{$\Delta i$ [\degree]}            & \multicolumn{4}{c}{$\Delta V_{tot}$  [m/s]}  \\ \cmidrule(lr){2-2} \cmidrule(lr){3-6} \cmidrule(lr){7-10}
			$\sigma_d$ [deg] &  Mean & Mean  & Min   & Max  & Std  & Mean    & Min     & Max      & Std   \\ \midrule
			105 &   0.038         & -0.023 & -0.081 & 0.082 & 0.033 & 92.29    & 59.98    & 130.61    & 11.05  \\
			120 &   0.036         & -0.023 & -0.096 & 0.111 & 0.034 & 74.41    & 55.35    & 114.77    & 8.12  \\
			135 &   0.034     & -0.018 & -0.818 & 0.207 & 0.050 & 64.41    & 49.13    & 137.25  & 7.93 \\ \bottomrule \bottomrule 
	\end{tabular}
\end{table}

\begin{figure}[tb!]\centering
	\captionsetup{justification=centering,margin=2cm}
    \includegraphics[trim={0 0.3cm 0 0},clip]{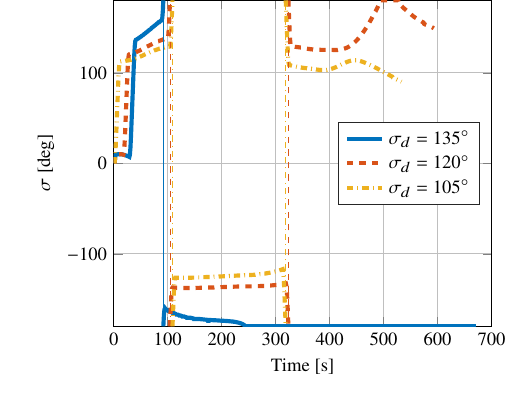}
	\caption{Bank angle history for OAK guidance with different values of $\sigma_d$, in a case of strong perturbations.}
	\label{fig:bank_angle_high_incl2}
\end{figure}

As a consequence, robustness should be evaluated in terms of $\Delta V_{tot}$, which is the $\Delta V$ required to correct orbit shape and plane at the same time. The analysis is limited for entry angles between -6\degree~and -5\degree{: for steeper entries, the controllability decreases, and the difference between different guidance schemes is smaller}. In this range, the optimal $\Delta V$ is approximately constant. Tables~\ref{tab:various_sigma_d_long} and \ref{tab:various_sigma_d_lat} summarize the main {results} for the OAK{1} guidance with different values of $\sigma_d$. On average, $\sigma_d$ = 135\degree ~provides the best results, both for in-plane and total $\Delta V$. Nonetheless, such setting causes a few cases in which the final inclination error is very large, as mentioned before. Saturation happens also for lower values of $\sigma_d$, but later and with much smaller effect in the final inclination error. Figure~\ref{fig:bank_angle_high_incl2} shows the bank angle history of the three guidance logics for same conditions and perturbations. Where the one with highest $\sigma_d$ saturates rapidly, leading to a final inclination error of more than 0.8\degree, the remaining two do not, leading to inclination errors of only 0.03\degree~($\sigma_d$ = 120\degree) and 0.01\degree~($\sigma_d$ = 105\degree).

\subsection{Comparison Between Different Rotational Models}
{In this subsection OAK1 and OAK2 are compared. To do so, the comparison needs to be carried out for equivalent values of $\sigma_d$. However, the interpretation of $\sigma_d$ varies between the to methods due to differences in their internal structures. As a result, a direct comparison at identical $\sigma_d$ values may be misleading. To enable a fair comparison, each OAK1 configuration is matched with the $\sigma_d$ value in OAK2 that yields the closest performance. To limit the scope of this search, we consider OAK1 at three representative values: 105\degree, 120\degree, 135\degree. The corresponding best-matching OAK2 values, identified at integer intervals, are 102\degree, 115\degree, and 128\degree, respectively. The mean absolute difference~(MAD) between the two algorithms is computed by averaging the absolute differences of their outputs over 1,000 Monte Carlo samples under identical conditions. Similarly, the mean absolute percentage difference~(MAPD) is computed by averaging the absolute relative differences. Table~\ref{tab:OAK1vs2} summarizes the performance comparison in terms of $\Delta V$ using MAD and MAPD. In all three cases, the MAD is less than 2.25~[m/s], and the MAPD is less than 0.25\%.
To put these numbers into perspective, the sample-wise difference in $\Delta V$ between flying OAK2 with $\sigma_d=115$\degree and $\sigma_d=114$\degree is equal to 0.9525~m/s. Hence, the difference between flying OAK1 and OAK2 is effectively equivalent to adjusting OAK2’s $\sigma_d$ by approximately one degree. From a practical standpoint, this suggests that the two rotational models exhibit comparable performance. Given its simpler structure, OAK1 is selected as the guidance logic for the remainder of this section.}
\begin{table}[]
	\centering
	\caption{Performance difference between OAK1 and OAK2 for comparable values of $\sigma_d$.}
	\label{tab:OAK1vs2}
		\begin{tabular}{cccccc}\toprule \toprule
		 \multicolumn{2}{c}{$\sigma_d$~[deg]}    &  \multicolumn{4}{c}{$|\Delta V_{OAK1}-\Delta V_{OAK2}|$}    \\   \cmidrule(lr){1-2}       \cmidrule(lr){3-6} 
			OAK1 & OAK2 & Mean~[m/s]  & Max~[m/s]   & MAD~[m/s] & MAPD   \\ \midrule
			105 & 102&  0.1112  & 18.7350       & 2.2091  & 0.24\% \\
			120 & 115&  0.3850  & 11.9556       & 1.6255  & 0.16\%  \\
			135 & 128 &  0.1982   & 12.2521  & 1.2986  & 0.12\%  \\ \bottomrule \bottomrule 
	\end{tabular}
\end{table}
\subsection{Sensitivity to vehicle parameters}
\begin{figure}[tb!]\centering
	\centering
	\captionsetup{justification=centering,margin=2cm}
	\captionsetup[subfigure]{justification=centering,singlelinecheck=false}
	\begin{subfigure}[t]{0.49\textwidth}\centering
        \includegraphics[trim={0 0.25cm 0 0},clip]{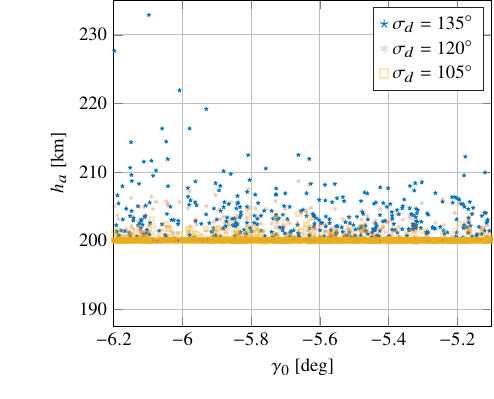}
		\caption{Apoapsis altitude.}
		\label{fig:Apollo_r_apo_105_120_135}
	\end{subfigure}
	\begin{subfigure}[t]{0.49\textwidth}\centering
        \includegraphics[trim={0 0.2cm 0 0},clip]{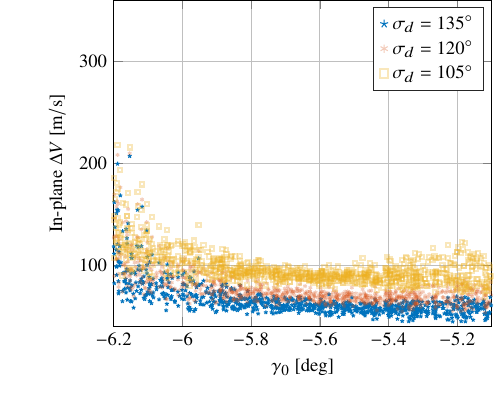}
		\caption{$\Delta V$.}
		\label{fig:Apollo_deltaV_105_120_135}
	\end{subfigure}
	\caption{Comparison between different values of $\sigma_d$ for the OAK1 guidance, flying Apollo CM.}
	\label{fig:Apollo_105_120_135}
\end{figure}

This subsection analyzes whether the same parameter tuning can work on different vehicles. The performance of {the OAK} guidance is analyzed using the Apollo {CSM} instead of the Orion {CM} as a reference vehicle. Apollo has a 20\% larger lift-to-drag ratio and a 7\% smaller ballistic coefficient. Further, the uncertainties in $C_D$ and $C_L$ both vary {by} $\pm20$\%, and independently from each other. Figure~\ref{fig:Apollo_105_120_135} shows the apoapsis altitude and planar $\Delta V$ for {the} Apollo {CSM} in a slightly different range of initial flight-path angles to reflect the difference in aerocapture corridor. The figure shows a pattern very similar to th{at} of Orion {CM}. The worst case of $\Delta V_{tot}$ is best for $\sigma_d$ = 120\degree, and equal to 96.7 m/s; the worst case with $\sigma_d$ = 135\degree~is 106.8 m/s instead.
Hence, the best $\sigma_d$ is approximately the same for both tested vehicles. 

\subsection{High-Speed Aerocapture}
This subsection shows results for higher speed aerocapture, with entry velocity {$V_0$ around} 16~km/s. The entry corridor is  adjusted accordingly, with $\gamma_0 \in [$-9.5\degree,- 6.5\degree]. Results are only compared with PredGuid+A Mode 6, since FNPAG would require new extensive tuning for these entry conditions.
\begin{figure}[tb!]
	\centering
	
        \includegraphics[trim={0 0.35cm 0 0},clip]{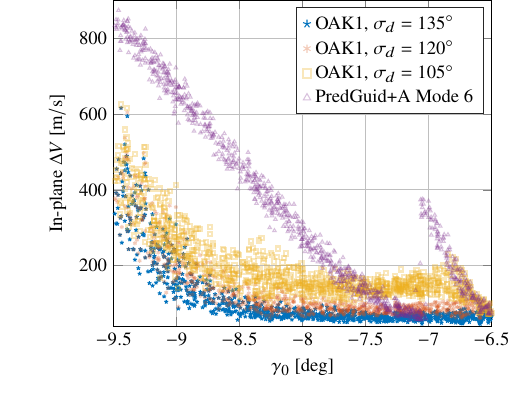}
	\caption{Comparison between PredGuid+A Mode 6 and OAK1 guidance, $V_0=$ 16~km/s.}
	\label{fig:V_16}
\end{figure}
Figure~\ref{fig:V_16} reports performance in terms of planar $\Delta V$. For {all} conditions under consideration, the OAK guidance consistently outperforms PredGuid+A Mode, as expected.
The results demonstrate again that the OAK guidance is easy to tune, since the same parameters for lunar return conditions give comparable results for higher speed aerocapture. Similarly to the slower aerocapture, there is a wide range of entry conditions where the OAK guidance can consistently provide a $\Delta V$ independently of initial flight path angle. The average of in-plane $\Delta V$ decreases with increasing $\sigma_d$ up to 135\degree. {As for the lower speed scenario, t}he optimal value of $\sigma_d$ is 120\degree ~when considering out-of-plane corrections, both in terms of worst case scenarios and average.

\subsection{Lateral Logic Performance}
\begin{figure}
    \centering
    \includegraphics[trim={0 0.5cm 0 0},clip]{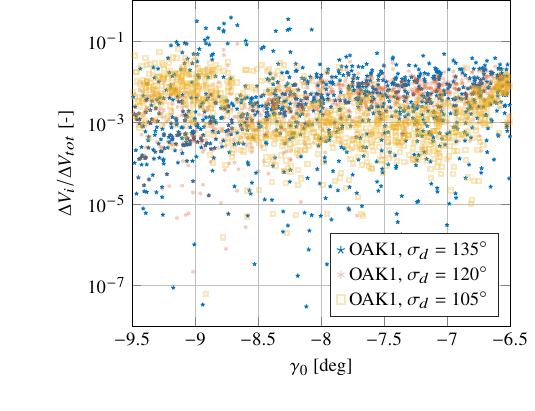}
    \caption{Lateral logic performance.}
    \label{fig:V_16_oop}
\end{figure}

The performance of the lateral guidance is evaluated for the high-speed aerocapture, with $V_0 = 16$~km/s. {Let $\Delta V_i$ be the difference between $\Delta V_{tot}$ and in-plane $\Delta V$. The inclination correction is performed simultaneously with the periapsis raise to save fuel. The performance metric for the lateral logic is then} the ratio between $\Delta V_i$, which is the difference , and the total $\Delta V_{tot}$, which includes periapsis raise, inclination correction, and apoapsis correction. Fig.~\ref{fig:V_16_oop} shows the results as a function of initial flight path angle and $\sigma_d$. As implied from previous results, the guidance {with} $\sigma_d=135$\degree ~induces unreliability in the lateral logic: the only cases where the inclination change accounts for more than 10\% {of the total $\Delta V$} occur with that setting. Moreover, steep entry angles are generally the cause for more inclination errors, which is to be expected because maneuverability is limited due keeping a bank angle very close to zero for a large portion of the trajectory.

\subsection{Heat Load Analysis}
This subsection validates the proof of Sec.~\ref{sub:min_rad_heatload_aero} in a {more realistic} environment.
For convective and radiative heat {fluxes}, a nose radius of 6.03 m has been used.
\begin{figure}[tb!]\centering
	\centering
    \captionsetup{justification=centering,margin=2cm}
	\captionsetup[subfigure]{justification=centering,singlelinecheck=false}
	\begin{subfigure}[t]{0.49\textwidth}\centering        \includegraphics{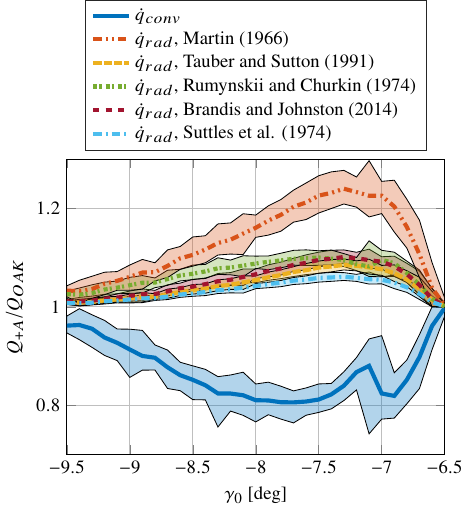}
		\caption{Radiative and convective, individually.}
		\label{fig:ratio_heatload}
	\end{subfigure}
	\begin{subfigure}[t]{0.49\textwidth}\centering		\includegraphics{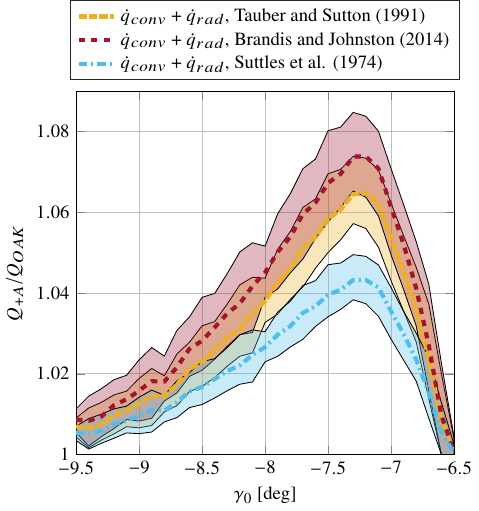}
    \caption{Total.}
		\label{fig:totalheat_V16}
	\end{subfigure}
	\caption{Ratio between total heat load with OAK1 guidance, $\sigma_d=$ 120\degree, and with PredGuid+A Mode 6.}
	\label{fig:ratios_heat}
\end{figure}
Figure~\ref{fig:ratios_heat} shows the ratio of different formulations of components of the heat loads between trajectories {with same initial conditions and perturbations} flown with OAK guidance and PredGuid+A Mode 6. {The figures were generated by grouping the results into bins of 0.05 degrees along the x-axis. For each bin, the minimum, mean, and maximum values were computed. The solid line shows the mean value in each bin, while the shaded region spans between the minimum and maximum values, connected across bin centers. As such, the shaded area approximates the range of the data and may not strictly contain 100\% of the samples.}
Figure~\ref{fig:ratio_heatload} is a numerical validation of Fig.~\ref{fig:heatloadratio_formulas}. Note that the ratios are less pronounced here, for two main reasons: 1) the OAK guidance does not follow an exact bang-bang trajectory because of the margin {$\sigma_d$}, {and} because of perturbations; and 2) the comparison is against a guidance that flies at constant bank angle, instead of flying lift down lift up {as in Fig.}~\ref{fig:heatloadratio_formulas}. As the convective heat {load} is {comparatively} much smaller in high-speed environments, a reduction by 5 to 10\% in radiative heat is more impactful on the total {heat load} than a reduction by 20\% in convective heat. This is further demonstrated by Fig.~\ref{fig:totalheat_V16}, which shows the ratio between the two guidance logics for the sums of convective and radiative heat {loads}. Independent of the formulations {used to compute the radiative heat flux, trajectories flown with OAK experience less total heat load than} the baseline. Some formulations of radiative heat flux have not been used for these results because of difficulties in recovering the proportionality constants.


\section{Conclusions}
\label{sec:conclusions}
This paper brings three major contributions. First, it is analytically proven that, for several analytical approximations of the radiative heat flux, the aerocapture trajectory minimizing the radiative heat load {is bang-bang, starting with lift up, and with a single switch}. The results are validated numerically. As radiative heat load grows rapidly with increasing velocity, its minimization is especially relevant for high-speed missions{, not only on Earth, but also, for example, on Mars, Titan, and Neptune}. Further, as radiative heat {flux} grows with the nose radius, whereas convective heat {flux} decreases {with the nose radius}, effect of radiation is more relevant for larger spacecraft. {For high speed aerocapture, radiative heat load can be reduced by 20 to 70\% compared to a trajectory flown with lift down lift up. Since it is proven that} the minimum radiative heat {load} trajectory coincides with the trajectory {that also} minimizes $\Delta V$, peak heat flux, and peak aerodynamic load, for missions where radiative heat load is much larger than convective heat load there is no need to seek trade-offs between costs {or constraints}.
The second contribution is the introduction of a new longitudinal guidance, which takes attitude constraints into account. This guidance achieves performance similar to the {current state-of-the-art} algorithms, but it does so {while requiring} less tuning. The tuning of the longitudinal guidance is robust to wide differences in initial flight path angle and velocity, as well as to different vehicles.
Third, a novel lateral guidance is introduced. The lateral guidance is robust and generally only requires one or two reversals. Few reversals lead to less interference with the longitudinal planni{n}g. {Empirical results show that} for lunar return conditions this lateral guidance guarantees a maximum inclination error of 0.082\degree, and the corresponding correction never exceeds 4 m/s. For high-speed aerocapture, the lateral correction is limited to less than 10\% of the total $\Delta V$.

\section*{Appendix}
{This appendix provides a proof for why lift up-lift down aerocapture trajectories maximize the altitude for every velocity. The proof is presented in two parts, assuming the motion is governed by the simplified equations~\eqref{eq:Vdot}-\eqref{eq:rdot}, with the additional assumption that $D/m\gg g|\sin\gamma|$, since $\gamma$ is small during aerocapture. The only assumption needed on the density is that it decreases monotonically with altitude. First, it is shown that a lift up trajectory results in higher altitude, for any given velocity, than any other trajectory originating from identical initial conditions. Second, it is demonstrated that a lift up-lift down trajectory designed to reach a specific apoapsis altitude attains a greater altitude, for any given velocity throughout its entire path, compared to any alternative trajectory with the same apoapsis target and initial conditions.}

{The derivatives of the radius $r$ and flight path angle $\gamma$ with respect to the velocity are, respectively}
\begin{equation}
\label{eq:drdv}
    {\frac{dr}{dV}= -2\frac{m\sin\gamma}{\rho(r) V S C_D},}
\end{equation}
\begin{equation}
\label{eq:dgdv}
    {\frac{d\gamma}{dV}=-\frac{C_L\cos\sigma}{C_D V} -\frac{2 m}{\rho(r) V^3 S C_D}\left( \frac{V^2}{r}-\frac{\mu}{r^2} \right) \cos\gamma.}
\end{equation}
{For a fixed value of $V$ and $r$, minimizing Eq.~\eqref{eq:drdv} (noting that minimization is desired since $\dot{V}$ is negative) requires increasing $\gamma$. Consequently, to ensure the altitude is maximized at future velocities, the trajectory should be designed to minimize~\eqref{eq:dgdv}. The minimization occurs when $\cos\sigma=1$.
Hence, an increasing $\gamma$, which results in a slower decrease (or more rapid increase) in $r$, also leads to a faster increase in $\gamma$. The influence of $\cos\gamma$ in the equation is neglected, since $\gamma$ is always close to zero. Note that the case of a skipping trajectory that later re-enters the atmosphere is not accounted for in this proof. During the coasting phase, the drag is almost zero, and thus, even with small flight path angle, the assumption that $D/m\gg g|\sin\gamma|$ is violated.}

{Based on the trajectory described above, the spacecraft will reach a point where the values of $r$, $V$, and $\gamma$ are such that the only way to achieve the desired apoapsis altitude is by flying the remainder of the trajectory full lift-down. As the radius is no longer maximized at each command, one might question whether a higher value of $r(V)$ could be achieved. However, for this to be possible, a smaller value of $\gamma$ would be required, allowing the spacecraft to still reach the desired apoapsis. This, however, would be impossible, since a lower value of $\gamma$ cannot lead to a larger value of $r$, per Eq.~\eqref{eq:drdv}.}

\section*{Acknowledgments}
Research was done at Delft University of Technology, in the Astrodynamics and Space Missions Section, when Enrico M. Zucchelli was an M.Sc. student.


\end{document}